\documentclass[showpacs,amsmath,amssymb,onecolumn,pra,superscriptaddress,notitlepage]{revtex4-2}

\usepackage[dvips]{graphicx}
\usepackage{subfigure}
\usepackage{amsfonts}
\usepackage{amssymb}
\usepackage{amscd}
\usepackage{amsmath}    
\usepackage{enumerate}
\usepackage{epsfig}
\usepackage{subfigure}
\usepackage{bm}
\usepackage{xcolor}
\usepackage{amsthm}
\usepackage{framed}
\usepackage{multirow}
\usepackage{mathrsfs,amssymb}
\usepackage{mathtools}
\usepackage{color}
\usepackage{longtable}
\usepackage{comment}
\usepackage[ruled,vlined]{algorithm2e}

\graphicspath{{./figure/}}

\begin{document}

\title{Implementation of a 46-node quantum metropolitan area network}

\author{Teng-Yun Chen}
\email{tychen@ustc.edu.cn}
\affiliation{Hefei National Laboratory for Physical Sciences at Microscale and Department of Modern Physics, University of Science and Technology of China, Hefei, Anhui 230026, China}
\affiliation{Shanghai Branch, CAS Center for Excellence and Synergetic Innovation Center in Quantum Information and Quantum Physics,
University of Science and Technology of China, Shanghai 201315, China}

\author{Xiao Jiang}
\affiliation{Hefei National Laboratory for Physical Sciences at Microscale and Department of Modern Physics, University of Science and Technology of China, Hefei, Anhui 230026, China}
\affiliation{Shanghai Branch, CAS Center for Excellence and Synergetic Innovation Center in Quantum Information and Quantum Physics,
University of Science and Technology of China, Shanghai 201315, China}

\author{Shi-Biao Tang}
\affiliation{QuantumCtek Co.~Ltd., Hefei, Anhui 230026, China}

\author{Lei Zhou}
\affiliation{QuantumCtek Co.~Ltd., Hefei, Anhui 230026, China}

\author{Xiao Yuan}
\affiliation{Center for Quantum Information, Institute for Interdisciplinary Information Sciences, Tsinghua University, Beijing 100084, China}

\author{Hongyi Zhou}
\affiliation{Center for Quantum Information, Institute for Interdisciplinary Information Sciences, Tsinghua University, Beijing 100084, China}

\author{Jian Wang}
\affiliation{Hefei National Laboratory for Physical Sciences at Microscale and Department of Modern Physics, University of Science and Technology of China, Hefei, Anhui 230026, China}
\affiliation{Shanghai Branch, CAS Center for Excellence and Synergetic Innovation Center in Quantum Information and Quantum Physics,
University of Science and Technology of China, Shanghai 201315, China}
\author{Yang Liu}
\affiliation{Hefei National Laboratory for Physical Sciences at Microscale and Department of Modern Physics, University of Science and Technology of China, Hefei, Anhui 230026, China}
\affiliation{Shanghai Branch, CAS Center for Excellence and Synergetic Innovation Center in Quantum Information and Quantum Physics,
University of Science and Technology of China, Shanghai 201315, China}

\author{Luo-Kan Chen}
\affiliation{Hefei National Laboratory for Physical Sciences at Microscale and Department of Modern Physics, University of Science and Technology of China, Hefei, Anhui 230026, China}
\affiliation{Shanghai Branch, CAS Center for Excellence and Synergetic Innovation Center in Quantum Information and Quantum Physics,
University of Science and Technology of China, Shanghai 201315, China}

\author{Wei-Yue Liu}
\affiliation{School of Information Science and Engineering, Ningbo University, Ningbo, Zhejiang 315211, China}

\author{Hong-Fei Zhang}
\affiliation{Hefei National Laboratory for Physical Sciences at Microscale and Department of Modern Physics, University of Science and Technology of China, Hefei, Anhui 230026, China}
\affiliation{Shanghai Branch, CAS Center for Excellence and Synergetic Innovation Center in Quantum Information and Quantum Physics,
University of Science and Technology of China, Shanghai 201315, China}

\author{Ke Cui}
\affiliation{Hefei National Laboratory for Physical Sciences at Microscale and Department of Modern Physics, University of Science and Technology of China, Hefei, Anhui 230026, China}
\affiliation{Shanghai Branch, CAS Center for Excellence and Synergetic Innovation Center in Quantum Information and Quantum Physics,
University of Science and Technology of China, Shanghai 201315, China}

\author{Hao Liang}
\affiliation{Hefei National Laboratory for Physical Sciences at Microscale and Department of Modern Physics, University of Science and Technology of China, Hefei, Anhui 230026, China}
\affiliation{Shanghai Branch, CAS Center for Excellence and Synergetic Innovation Center in Quantum Information and Quantum Physics,
University of Science and Technology of China, Shanghai 201315, China}

\author{Xiao-Gang Li}
\affiliation{QuantumCtek Co.~Ltd., Hefei, Anhui 230026, China}

\author{Yingqiu Mao}
\affiliation{Hefei National Laboratory for Physical Sciences at Microscale and Department of Modern Physics, University of Science and Technology of China, Hefei, Anhui 230026, China}
\affiliation{Shanghai Branch, CAS Center for Excellence and Synergetic Innovation Center in Quantum Information and Quantum Physics,
University of Science and Technology of China, Shanghai 201315, China}

\author{Liu-Jun Wang}
\affiliation{Hefei National Laboratory for Physical Sciences at Microscale and Department of Modern Physics, University of Science and Technology of China, Hefei, Anhui 230026, China}
\affiliation{Shanghai Branch, CAS Center for Excellence and Synergetic Innovation Center in Quantum Information and Quantum Physics,
University of Science and Technology of China, Shanghai 201315, China}

\author{Si-Bo Feng}
\affiliation{QuantumCtek Co.~Ltd., Hefei, Anhui 230026, China}

\author{Qing Chen}
\affiliation{QuantumCtek Co.~Ltd., Hefei, Anhui 230026, China}

\author{Qiang Zhang}
\affiliation{Hefei National Laboratory for Physical Sciences at Microscale and Department of Modern Physics, University of Science and Technology of China, Hefei, Anhui 230026, China}
\affiliation{Shanghai Branch, CAS Center for Excellence and Synergetic Innovation Center in Quantum Information and Quantum Physics,
University of Science and Technology of China, Shanghai 201315, China}

\author{Li Li}
\affiliation{Hefei National Laboratory for Physical Sciences at Microscale and Department of Modern Physics, University of Science and Technology of China, Hefei, Anhui 230026, China}
\affiliation{Shanghai Branch, CAS Center for Excellence and Synergetic Innovation Center in Quantum Information and Quantum Physics,
University of Science and Technology of China, Shanghai 201315, China}

\author{Nai-Le Liu}
\affiliation{Hefei National Laboratory for Physical Sciences at Microscale and Department of Modern Physics, University of Science and Technology of China, Hefei, Anhui 230026, China}
\affiliation{Shanghai Branch, CAS Center for Excellence and Synergetic Innovation Center in Quantum Information and Quantum Physics,
University of Science and Technology of China, Shanghai 201315, China}

\author{Cheng-Zhi Peng}
\affiliation{Hefei National Laboratory for Physical Sciences at Microscale and Department of Modern Physics, University of Science and Technology of China, Hefei, Anhui 230026, China}
\affiliation{Shanghai Branch, CAS Center for Excellence and Synergetic Innovation Center in Quantum Information and Quantum Physics,
University of Science and Technology of China, Shanghai 201315, China}

\author{Xiongfeng Ma}
\affiliation{Center for Quantum Information, Institute for Interdisciplinary Information Sciences, Tsinghua University, Beijing 100084, China}

\author{Yong Zhao}
\affiliation{Hefei National Laboratory for Physical Sciences at Microscale and Department of Modern Physics, University of Science and Technology of China, Hefei, Anhui 230026, China}
\affiliation{Shanghai Branch, CAS Center for Excellence and Synergetic Innovation Center in Quantum Information and Quantum Physics,
University of Science and Technology of China, Shanghai 201315, China}
\affiliation{QuantumCtek Co.~Ltd., Hefei, Anhui 230026, China}

\author{Jian-Wei Pan}
\affiliation{Hefei National Laboratory for Physical Sciences at Microscale and Department of Modern Physics, University of Science and Technology of China, Hefei, Anhui 230026, China}
\affiliation{Shanghai Branch, CAS Center for Excellence and Synergetic Innovation Center in Quantum Information and Quantum Physics,
University of Science and Technology of China, Shanghai 201315, China}



\begin{abstract}
Quantum key distribution (QKD) enables secure key exchanges between two remote users. The ultimate goal of secure communication is to establish a global quantum network. The existing field tests suggest that quantum networks are feasible. To achieve a practical quantum network, we need to overcome several challenges, including realising versatile topologies for large scales, simple network maintenance, extendable configuration, and robustness to node failures. To this end, we present a field operation of a quantum metropolitan-area network with 46 nodes and show that all these challenges can be overcome with cutting-edge quantum technologies. In particular, we realise different topological structures and continuously run the network for 31 months, by employing standard equipment for network maintenance with an extendable configuration. We realise QKD pairing and key management with a sophisticated key control center. In this implementation, the final keys have been used for secure communication such as real-time voice telephone, text messaging, and file transmission with one-time pad encryption, which can support 11 pairs of users to make audio calls simultaneously. Combined with inter-city quantum backbone and ground-satellite links, our metropolitan implementation paves the way toward a global quantum network.
\end{abstract}

\maketitle

\section*{Introduction}
The ultimate goal of quantum key distribution (QKD) \cite{BB84,Ekert:QKD:1991} is to construct a global quantum network, wherein all communication traffics have information-theoretical security guarantees. A global QKD network consists of two main types of links: the ground network (mainly fibre based) and the satellite network (mainly free-space based). The ground network can be further divided into backbone, metropolitan, and access networks, which cover inter-city distances, metropolitan distances, and fibre-to-the-home distances, respectively. The feasibility of QKD between two users has been extensively studied, for example, through long-distance free space \cite{Ursin07}, telecom fibres \cite{Takesue07}, and simulated ground-satellite links \cite{nauerth2013air,Wang13}. Field tests of QKD networks have been realised, including the three-user network by DARPA (2003) \cite{elliott2005current}, the six-node SECOQC network in Europe (2008) \cite{peev2009secoqc}, SwissQuantum network (2009) \cite{Stucki2011}, the USTC network \cite{Wang2010wavelengthsaving}, the six-node mesh-type network in Tokyo (2011) \cite{Sasaki11}, and the small-scale metropolitan all-pass and inter-city quantum network \cite{chen2009field,Chen10}. The satellite network is a promising way to realise intercontinental secure communication due to the low transmission attenuation in space. The satellite can serve as a trusted relay, connecting remote user nodes or sub-networks \cite{PhysRevLett.120.030501}. Recently, a large-scale satellite network has been implemented \cite{chen2021integrated}, consisting of four metropolitan-area networks, a backbone network and two satellite–ground links. Here, we summarize the existing network implementations in Table~\ref{tab:comparison}. For a full review of the subject, one can refer to the recent review article \cite{Xu2020Secure} and references therein.

\begin{table*}[hbt]
\centering
\footnotesize
\caption{Existing QKD network implementations.} \label{tab:comparison}
\begin{tabular}{l @{\hspace{0.3cm}} l  @{\hspace{0.3cm}} l @{\hspace{0.3cm}} l @{\hspace{0.3cm}} l  }
  \hline \hline
      \textbf{Network} & \textbf{\# of Nodes} & \textbf{Running time} & \textbf{Topology} & \textbf{Key rate }
\tabularnewline
&                 &{\small \textbf{(order of magnitude)}} &          &{\small \textbf{(at max distance/loss) }}    \\
   \hline
   DARPA~ \cite{elliott2005current} & $6$ & unknown & tree & $0.5$ kbps ($10.2$ km)  \\
 SECOQC~ \cite{peev2009secoqc}& $6$ & hour & mesh & $3.1$ kbps ($33$ km) \\
   SwissQuantum~ \cite{Stucki2011}& $3$ & year & fully-connected & $1$ kbps ($17.1$ km)\\
   Tokyo~ \cite{Sasaki11}& $6$ & year & mesh & $2.2$ kbps ($90$ km)\\
    USTC~ \cite{Wang2010wavelengthsaving}& $5$ & unknown & star & $0.4$ kbps ($14.8$ dB)\\
    USTC~ \cite{chen2009field} & $3$ & week & fully-connected & $1.5$ kbps ($20$ km) \\
    USTC~ \cite{Chen10}& $5$ & week & star & $0.2$ kbps ($130$ km) \\
    USTC~  \cite{chen2021integrated} & $109$& year & mesh & 1.1kbps ($2043$ km) \\
    Hefei (This work)& $46$ & year & fully-connected \& star & $49.5$ kbps ($18$ km)\\
   \hline \hline
\end{tabular}
\end{table*}

Nevertheless, these QKD experiments and networks are still preliminary demonstrations with limited scales with less than ten nodes, making it insufficient for meeting the demands of actual metropolitan communication. Furthermore, realising a practical QKD network is not simply extending the number of nodes; while many scientific and practical issues, such as (a) network topology, (b) network scalability, (c) key management, (d) practical applications, and (e) network robustness, need to be considered. Thus far, realising a practical large QKD network still remains as a major challenge in quantum communication.

\begin{figure*}[hbt]
\centering
\includegraphics[width=12cm]{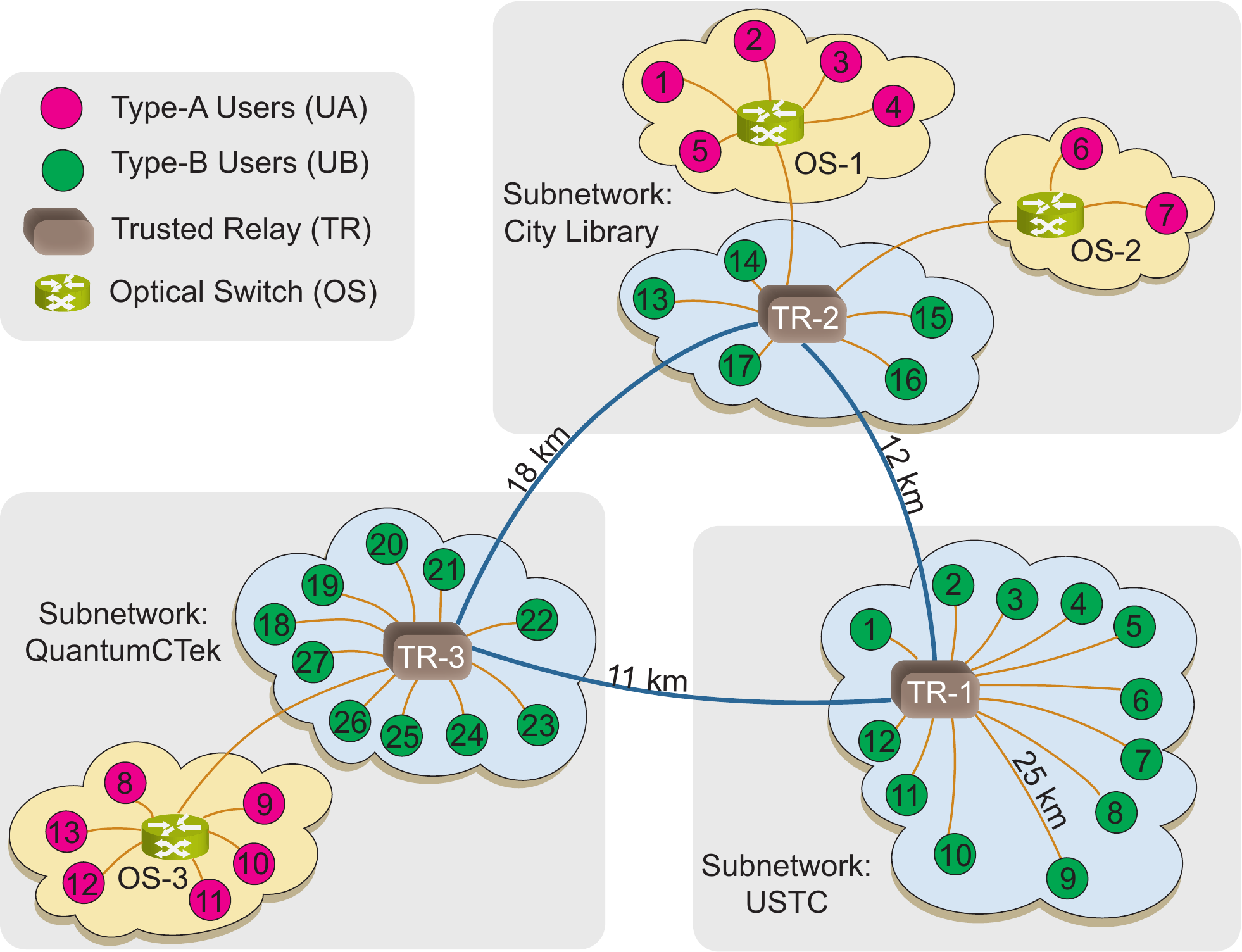}
\caption{The topological structure of our quantum network. The network mainly comprises three sub-networks that are directly connected to each other. In each sub-network, there are multiple users connected to intermediate nodes in different ways, either by an all-pass optical switch (OS) or a trusted relay (TR). Users connected by a switch are denoted as red dots (Type-A Users, UA), holding both a quantum transmitter and a receiver. Users connected to a trusted relay are denoted as green dots (Type-B Users, UB), only holding a quantum transmitter. Specifically, UA-1 to UA-5 are connected to OS-1, UA-6 and UA-7 are connected to OS-2, UA-8 to UA-13 are connected to OS-3, UB-1 to UB-12 are connected to TR-1, UB-13 to UB-17 are connected to TR-2, and UB-18 to UB-27 are connected to TR-3.} \label{Fig:Net}
\end{figure*}

In this work, we construct a 46-node quantum metropolitan area network throughout the city of Hefei, which connects 40 user nodes, three trusted relays, and three optical switches, as shown in Fig.~\ref{Fig:Net}. The network covers the entire urban area and connects several major organizations in the city districts, including governments, banks, hospitals, universities, and research institutes. In our network, we (a) implement versatile connection topologies for different hierarchies of users, (b) use standard equipments with a scalable configuration, (c) integrate systematic key management, (d) realise various robust application modules, and (e) deal with node failures. As a result, we address the major challenges in realising a large scale practical QKD network.

\section*{Results}
\subsection*{Network topology}
We first review the basic topological structures in a network. There are three general ways of connecting and distributing keys between users in a quantum network. The most robust method uses a fully connected topology. Here, each user is directly connected to every other user in the network. This type of network contains no relays; hence it is robust against a single point of failure, and the users do not need to trust one another. That is to say that a system failure or dishonest user would not affect the communication between other users. The main drawback of this type of network is that the number of links (and cost) of a fully connected network quadratically increases with the number of users. Thus, such a network is typically used for connections between a small number of major nodes.

Alternatively, the user nodes can also be connected via a central switch (relay). In this star-like network, the number of links linearly increases with the number of users. In addition, the users do not need to trust each other or the relay. Because the switch only transfers quantum signals, users can execute QKD protocols as if they are directly connected. The drawback of this type of network is that it is not robust against a single point of failure. That is to say that if the switch relay fails, the entire network will be brought down. The transmission distance of quantum signals is twice the length of the link between the users and the switch; hence this kind of network is typically used for local connections.

In the star-like topology, we can replace the switch with a trusted node. In this trusted node network, every user runs QKD protocols with a central relay, and two users can combine their keys between the central relay to form their own keys. In QKD, the secure key transmission distance is limited; thus, the size of a directly communicated quantum network is also limited. However, the size of the network can be extended by the introduction of trusted relays. Two distant users could also build secure keys with the help of a sufficient number of trusted relays. In practice, the Shanghai--Beijing backbone employs this technique to scale the QKD distance. The disadvantage of this type of network is that the users need to trust the relay. To construct a global quantum network, it is important to realize different topological structures in practice.

Our network consists of three sub-networks located at, USTC, QuantumCTek, and the City Library, and are distributed approximately 15 kms apart.
The longest fibres connecting the east and west end-users is approximately 45 km, and that connecting the south and north end-users is approximately 42 km. The longest direct distance between two users in the network is approximately $18$ km.
We realise two basic types of topological connection structures, including the full connection between the three subnetworks and the star-like connection for local access networks. The fully connected topology is applied to guarantee the robustness between the most important users; while the star-like connection is used for a more efficient network connection.
At the center of the star-like subnetwork, we either use a trusted node or an optical switch for different scenarios depending on the needs and distribution of the users.

The trusted node can be regarded as a classical router that assigns classical keys between users.
The all-pass optical switches acted as quantum routers that redistribute quantum signals. Any two users connected to the same switch could communicate directly without interfering with other users. In the experiment, we made use of two types of optical switches.
One is the $4\times8$ switch where four $1\times8$ optical switch modules and eight $1\times4$ modules are connected. This type of switch module comprises $4$ input and $8$ output ports, forming a $4\times8$ connecting matrix. The other is the $16$-port all-pass optical switch where sixteen $1\times15$ optical switch modules are connected to form an optical path. When this $16$-port switch was fully connected, it enables 8 pairs of users to communicate simultaneously.
In our experiments, the losses of all these optical switches are below $1.2$ dB, which are much lower than the channel isolation ($50$ dB).

\subsection*{Standard QKD equipments}
In our network, we used the polarization-encoding BB84 QKD protocol \cite{Hwang03,Lo05,Wang05} with a vacuum+weak decoy-state method \cite{Ma05} to generate secrete keys between directly connected users and trusted relays. Two users could generate keys if one of them had a quantum transmitter and the other had a quantum receiver. As a quantum receiver is generally more expensive compared with a quantum transmitter, not all users in this network possessed quantum receivers. However, everyone at least had a quantum transmitter and was thus able to transmit signals. In this case, there were two types of users in this network: users directly connected to a switch have both quantum a transmitter and receiver, and users directly connected to a trusted relay have only a quantum transmitter. There were, correspondingly, two types of equipment: one only for transmitting signals and the other for transmitting and receiving signals at the same time.

\begin{figure*}[hbt]
\centering
\includegraphics[width=12cm]{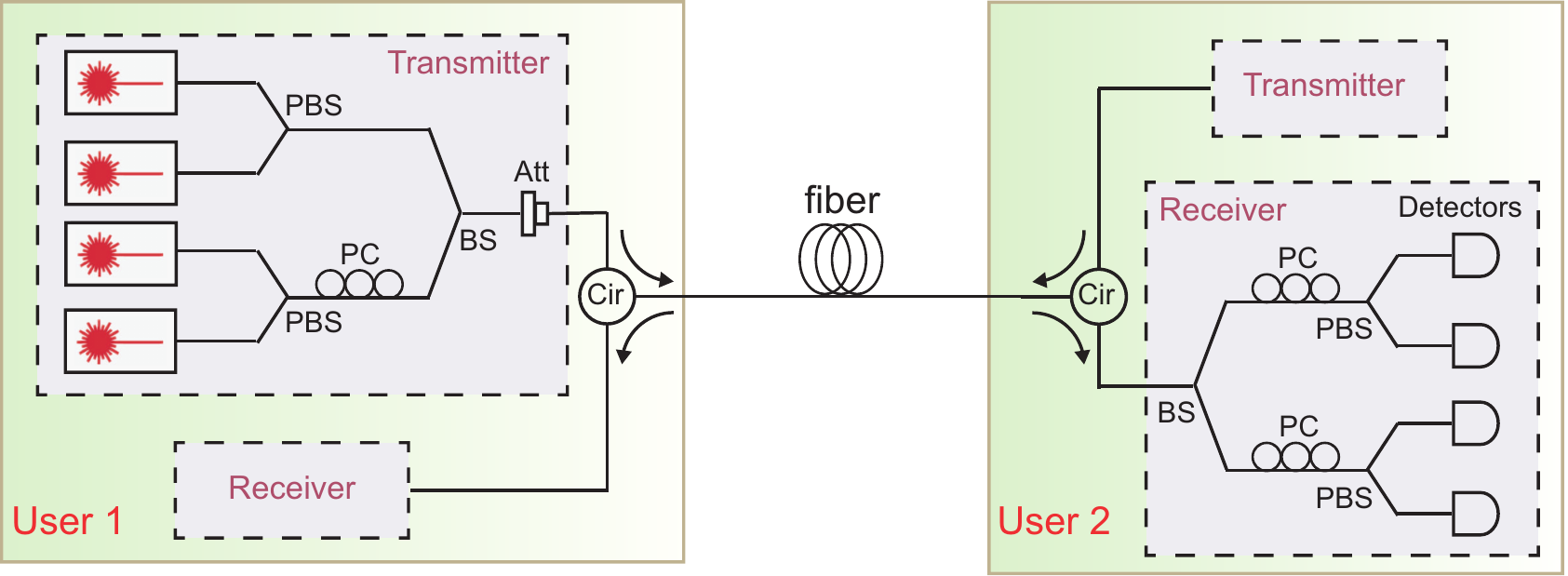}
\caption{A schematic for the QKD set-up. There are four laser sources in the transmitter emitting four corresponding polarisation states in the BB84 protocol. The polarisation is modulated via the PBS and the PC, and the average light intensity is modulated via the attenuator. Each laser produces three light pulses with different intensities including signal, decoy and vacuum states. The signal and decoy states contain mean photon numbers of $0.6$ and $0.2$ per pulse, respectively, and the ratio between the signal, decoy, and vacuum states is $6:1:1$. The optical misalignment is less than $0.5\%$. In the detection side, a four-channel InGaAs single photon detector is integrated with the following parameters. The detection efficiency is $10\%$, the dark count is $10^{-6}$, the dead time is $2$ $\mu$s, the afterpulse probability is less than $0.5\%$, and the effective gate width is $500$ ps. The receiver detects the light signal with the PC as a polarisation feedback. The Cir is used to realise transmission and reception of light signals simultaneously.
BS: beam splitter; PBS: polarising beam splitter; PC: polarisation controller; Att: attenuator; Cir: circulator.} \label{Fig:Equip}
\end{figure*}

Standard transmitter and receivers are applied in our network, whose internal structures are shown in Fig.~\ref{Fig:Equip}. In the transmitters, we use the $14$-pin butterfly distribute feedback lasers with central wavelength $1550$ nm. Polarization states \{$\left| H \right\rangle$, $\left| V \right\rangle$, $\left| + \right\rangle$, $\left| - \right\rangle$\} are produced with four different lasers, where each one can produce three different intensity pulses corresponding to the signal, decoy, and vacuum states. Before key generation, a time calibration between the source and the single-photon detectors as well as polarization feedback is performed. In general, the calibration is more efficient with strong pulses. It will take more time to complete the calibration for longer transmission distance but no more than 5 minutes. The calibration process makes our network robust against environmental disturbances. After basis reconciliation and error correction, privacy amplification is performed after $256$ kbit per second (kbps) keys are accumulated. Based on a field-programmable gate array (FPGA), the Winnow algorithm \cite{Buttler2003} is used for error correction, with a correction efficiency of $1.3\sim1.5$. Then, privacy amplification is performed using an FPGA implemented Toeplitz matrix Hash operation \cite{krawczyk1995new}, which is constructed by true random numbers shared by the transmitter and receiver devices. The standardization of the QKD equipment can greatly reduce the amount of devices required:  allowing the number of devices scales linearly with the number of user nodes.

\subsection*{Key management}
A key management strategy enables the users whose keys are running out to generate keys in high priority. We realise systematic key management for our network by designing a switching strategy. The strategy is determined by the amount of keys stored in the local memories for the users. The user with the least key amount has priority in the queue for key distribution. Here we take the $16$-port all-pass optical switch mentioned above as an example. Since it can be connected to $16$ users, there are a total of $\binom{16}{2}=120$ possible key pairing schemes by which two users are connected for the following QKD process. The queuing process for the key pairing scheme is determined by the Roll-Call-Polling protocol that judges the amount of keys between users. When the key amounts of all devices are the same, QKD pairing is sequentially performed in the order of the network ID. For arbitrary communication partners, the latency for key pairing is heuristically set to be $10$, $15$ or $30$ minutes according to experience.
Then the optical paths of the optical switch are connected, and the QKD process begins. Such a pairing process will repeat whenever there are QKD tasks. The switching time can be configured, ranging from $10$ to $60$ minutes. If two users in different subnetworks wish to perform QKD, they first generate keys with intermediate nodes and then swap them. After key generation is activated, the user can obtain secure keys within $5$ minutes, which are stored in local memories.

Since our network is scalable, we also need to consider the key management for new users. To join the network, a new user should first send a heartbeat frame from their QKD device to the key management server, i.e., its upstream optical switch or trusted relay node. A sequence of $32$ kbit initial keys with the trusted relay or optical switch is used for authentication. The authentication is implemented by the HMAC algorithm based on the symmetric key algorithm SM4 \cite{krawczyk1997hmac}, which does not provide information-theoretic security. Within $2$ minutes after power-on, the QKD device is connected to the network. Then, the device is in the queue for key generation.

\subsection*{Security analysis}
We follow the standard decoy state BB84 security analysis \cite{Lo05,Ma05} and its finite size analysis \cite{PhysRevA.81.012318}. The secret key rate of the BB84 protocol is given by \cite{gottesman2004security,Lo05},
\begin{equation}\label{eq:keyrateBB84}
r=-fQ_\mu H(E_\mu)+Y_1\mu e^{-\mu}[1-H(e_1^p)],
\end{equation}
where $f$ is the error correction efficiency, $\mu$ is the mean photon number per pulse for a signal state, $Q_\mu$ is the overall gain for the signal states, $E_\mu$ is the quantum bit error rate (QBER), $Y_1$, $e_1^p$ are the yield and phase error rate of the single photon component and $H(p)=-p\log_2 p-(1-p)\log_2(1-p)$ is the binary Shannon entropy function. The single photon yield and phase error rate can be well estimated by the decoy state method \cite{Lo05}. In fact, only three intensities (signal, weak decoy and vacuum) are enough to give tight bounds \cite{Ma05}, as implemented in our network.

In the finite size case, there will be deviations in the estimations of parameters above due to the statistical fluctuations. The main finite-size effect comes from the phase error rate estimation \cite{PhysRevA.81.012318}. Suppose we use $Z$-basis states to generate key, then the single photon phase error rate in this basis $e_1^{pz}$ is bounded by the single photon bit error rate in $X$ basis $e_1^{bx}$ and a small deviation $\theta$ optimized according to experimental data \cite{PhysRevA.81.012318},
\begin{equation}
e_1^{pz}\leq e_1^{bx}+\theta
\end{equation}
with a failure probability of
\begin{equation}
\epsilon_{ph}\leq \frac{\sqrt{n_x+n_z}}{\sqrt{e_1^{bx}(1-e_1^{bx})n_xn_z}}2^{-(n_x+n_z)\xi(\theta)}
\end{equation}
where $n_x$ and $n_z$ are the numbers of bits measured in $X$ and $Z$ basis, respectively, and $\xi(\theta)=H(e_1^{bx}+\theta-n_x \theta/(n_x+n_z))-n_xH(e_1^{bx})/(n_x+n_z)-(1-n_x/(n_x+n_z))H(e_1^{bx}+\theta)$. There will also be failure probabilities in other steps including the authentication, error verification, and privacy amplification. These failure probabilities are functions of the secure key consumption in the corresponding steps, and have additivity due to the composable security. In the Supplementary Note 2, we will show how to calculate the finite-size key rate in detail.

\subsection*{Application}
For applications of our network, users could make use of the generated secure keys to confidentially transfer information. The message is encoded in FPGA modules with an exclusive or operation on the secure keys. We apply our network to transmit encrypted information such as real-time voice telephone, instant messaging, and digital files with the one-time pad encryption method \cite{shannon1949communication}. The total amount of information to be encrypted is $10$ Gbits. The encryption speed is $800$ Mbps. The total delay in the encryption process is less than $50$ us.
In our network, the speed of real-time voice telephone was $2.4$ kbps, the speed of file transmission was 320 kbps. The capacity of our network is tested for $50$ minutes, as shown in Fig.~\ref{Fig:Res}. 22 users simultaneously made calls in the quantum network for $6$ minutes (see Supplementary Note 1 for more details).
\begin{figure*}[hbt]
\centering
\includegraphics[width=12cm]{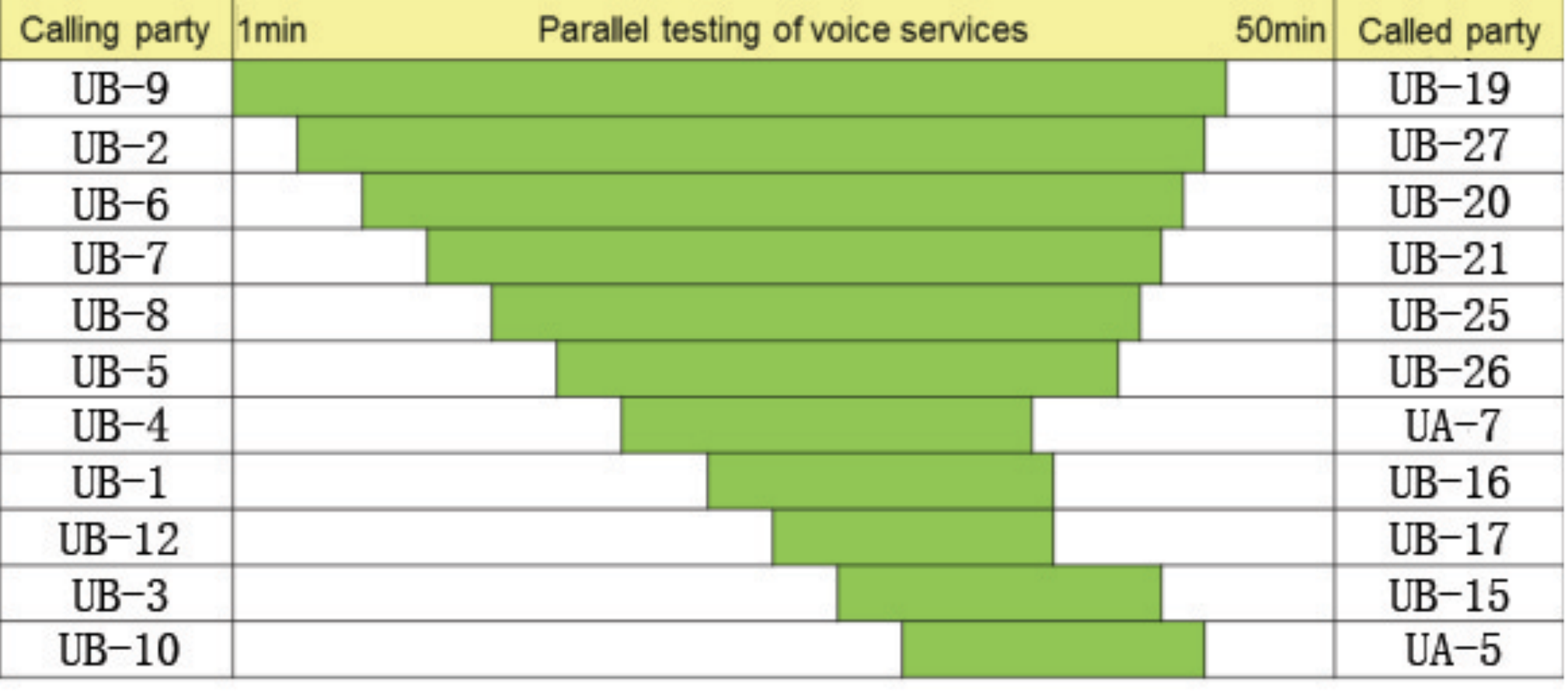}
\caption{Twenty-two users simultaneously make calls with QKD protocols. The green areas represent the duration over which users make calls.} \label{Fig:Res}
\end{figure*}

\subsection*{Network robustness}
\begin{figure*}[hbt]
\centering
\includegraphics[width=12cm]{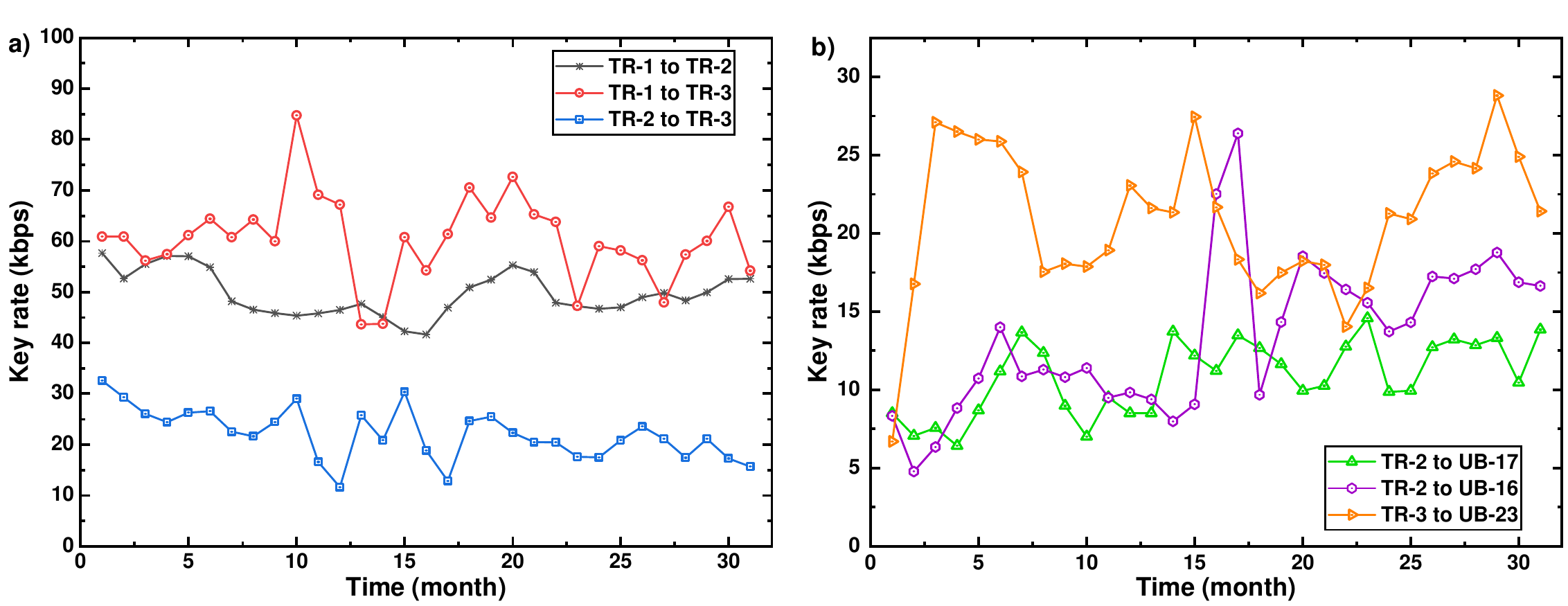}
\caption{The key rates versus time for some representative links. a) The key rates between the three trusted relays. b) The key rates between trusted relay and user. In the robustness test, $11$ user nodes have continuously run for $31$ months. The key rates are recorded every 30 seconds and taken average over a month. The detailed key rates are given as Supplementary Tables V and VI.}
\label{fig:keyratevstime}
\end{figure*}

In addition, the stability and robustness of the network were tested by running continuously for 31 months. We choose some representative nodes and show the key rates versus time in Fig.~\ref{fig:keyratevstime}. The key rate results are summarized in Table~\ref{tab:summary}, ranging from $6$ to $60.5$ kbps. Since the Hefei network is based on the Roll-Call-Polling protocol, the results are all average key rates during the QKD process. The key rate fluctuation mainly comes from the fast variations of photon polarization, which is determined by the internal structure and surrounding environment of the optical fiber. The error rate caused by the variations of photon polarization will accumulate with the propagation of the photons, leading to a drop of the key rate. Once the error rate is high enough, the QKD process is aborted and calibration is performed. Then the key rate will return to a normal value, corresponding to the ascensions in key rate performance.

\begin{table*}[ht!]
\centering
\caption{List of the average key rates between subnetworks and the key rate ranges with in the three subnetworks (lower). The detailed key rates are presented in the Supplementary Tables II, III and IV.}
\begin{tabular}{cccc}
\hline
\hline
 & $\textbf{TR1-TR2}$ & $\textbf{TR1-TR3}$ & $\textbf{TR2-TR3}$\tabularnewline
\hline
\textbf{Average key rate (kbps)} & $22.1$ & $49.7$ & $60.5$\tabularnewline\hline
\hline
 & \textbf{TR1 Subnetwork} & \textbf{TR2 Subnetwork} & \textbf{TR3 Subnetwork}\tabularnewline
\hline
\textbf{Key rate range (kbps)}& $6\sim17$ & $10\sim30$ & $6\sim37$\tabularnewline
\hline
\hline
\end{tabular}
\label{tab:summary}
\end{table*}

\section*{Discussion}
In summary, we have presented a practical, large-scale metropolitan QKD network with standard commercial QKD products, systematic key management, and practical usage in Hefei, China. This quantum network can be scaled by adding more users and relays, and it can be connected to the Shanghai--Beijing backbone to become a national network.
Our network can be combined with other QKD protocols that are robust against device imperfections. For instance, to overcome the imperfection of measurement devices, measurement-device-independent (MDI) QKD protocols \cite{Lo12} can be employed. In experiment, the MDI-QKD protocol has been extensively verified and an MDI-QKD network over unreliable metropolitan has been recently realized \cite{Tang16}.
Combined with the MDI-QKD network, one can imagine that communication in the future can be done in both efficient and secure way. Recently  an inter-continental QKD network was reported \cite{chen2021integrated}, connecting several metropolitan networks with a satellite. Our practical implementations and applications of a metropolitan network can be well combined with \cite{chen2021integrated} for future directions.



\section*{Acknowledgments}
The authors acknowledge insightful discussions with Z.~Zhang. This work was supported by the National Key Research and Development Program of China (2017YFA0303903), Anhui Initiative in Quantum Information Technologies, and the Chinese Academy of Sciences. The authors also acknowledge support from the Anhui Provincial Government, the Hefei City Government, the Hefei Broadcast \& TV Broad Band Network Ltd.

%

\section*{Supplementary Information}
\subsection*{Supplementary Note 1: Experiment Details}\label{app:detail}
\subsubsection{Details of the experimental parameters}
Due to the finite polarization contrast and the dark counts of single photon detectors, a sifted key can differ between Alice and Bob after basis comparison. These keys must be reconciled using classical error correction. In a QKD system, we use a field-programmable gate array (FPGA) based Winnow error correction algorithm with efficiency of $1.3\sim1.5$, and the reconciliation takes less than $50$ ms for a sifted key of $256$ kbits, when the bit error rate is less than $4$\%. After error correction, we apply a cyclic redundancy check (CRC). If the CRC passes, the key string is retained; otherwise, it is abandoned.

We use lasers as a light source in this system. However, doing so results in multi-photon constituents that could be used by eavesdroppers. Fortunately, we can reduce the amount of leaked information via privacy amplification. We estimate the secure factor using the decoy method. Privacy amplification is completely implemented in FPGA by a Toeplitz matrix. During the matrix multiplication process, we use block operations because the matrix is very large. The large matrix is divided into many smaller blocks; we can multiply two blocks in one clock cycle using parallel arithmetic. Note that the finite key effect is considered in the parameter estimation.

With the aid of automated monitoring and control, our system can run consistently for a long time. For instance, at a fibre length of $20$ km, the quantum bit error rate is generally below $1$\%, and the QKD systems can reach a key rate of $30$ kbps when running in both directions. The clock rate of our system is 20 MHz and is updated to $40$ MHz for the robustness test. The total loss includes that caused by the fibres, which is proportional to one fourth of their length in units of kilometers, as well as the inherent loss of the detector, which is about $3$ dB, and the efficiency of the detectors, which is about $10$\%.

\subsubsection{Details of the calibration}
The calibration of the QKD systems is performed in three main steps. Here all the optical pulses are strong pulses and not at single-photon level. The first step is associated with detector gate time calibration, where Alice's lasers sends optical pulses of $40$ MHz, and when the photon count rate, $D_1$, $D_2$, $D_3$, and $D_4$, of the four single photon detectors reach the maximum simultaneously, a visibility $|D_i-D_j |/|D_i+D_j|$  ($i,j=1,2,3,4$) is calculated. The gate positions of the detectors are scanned to maximize this visibility. The lowest value allowed for the QKD systems in this network is $15$\%. If the visibility is lower than this value for 3 times in a row for any detector, the calibration is aborted.

The second step is associated with polarization feedback. Specifically, it begins with Alice's sending $H$ polarization laser pulses at $40$ MHz to Bob. Bob adjusts his electrical polarization controllers to maximize the visibility. The minimum value of the polarization visibility permitted by the systems in the Hefei network is set to be $20$ dB. If the measured value is lower than this value, the electrical polarization controllers are initialized and scanned again. Once the polarization calibration for $H$ passes, Alice immediately switches the $H$ laser off and send $V$ pulses to verify the latter’s polarization visibility, which should be around the same value as $H$. The process is repeated for the $+$ and $-$ polarization pulses, and the calibration is aborted after 3 failed trials.

In the final step the repetition of Alice’s lasers are adjusted to $100$ kHz to match the synchronization laser pulses ($100$ kHz, wavelength at $1570$ nm). The four lasers send pulses to Bob to calibrate the time sequence. If any of the lasers fail to pass the synchronization after 3 consecutive times, the calibration is aborted.

The calibration performs better with stronger pulses. Therefore, when the transmission distance increases, the QKD system needs to increase the laser power. This makes the total calibration time for the three steps vary with the transmission distance. It turn out that the total time is always within 5 minutes.

The main source of noise in our network comes from the slow drifts in field fibers that affects the polarization of the quantum signals. our log file about the feedback calibration time can reflect how fast the noise builds up. For instance, over 24 hours on April 23, 2021, polarization of the TR2-TR3 link was calibrated for $15$ times, as shown Table below. As we can see, the polarization was very stable overnight and the QKD system can run steadily for almost $11$ hours. Over the daytime, noise was more prominent and polarization was calibrated every $56$ minutes.

\subsubsection{Details of the key management}
The network topology is pre-configured on the control center of the network. When each key management device is activated, they will log in the control center for authentication. The control center will control the devices that have already logged in and been authenticated to perform the QKD process and provide relevant necessary information, such as the connection configuration of neighboring devices.

The key will be automatically supplemented for authentication after the consumption. If the keys are lost for some reasons, one needs to manually reset the key. The authentication keys are stored separately, which are used only for identity authentication and cannot be accessed by other services.

We have employed several means to ensure security of the key management devices. For key storage, we encrypt the new secure keys with previously generated keys at the key pools and then store the encrypted keys in the secure modules. With FPGA-based protocols for isolation, these modules are protected against external network attacks. We apply anti-detection design and restrict the electromagnetic radiations.

\subsubsection{Details of the key rates}
We list the detailed key rates of the Hefei quantum key distribution network in Supplementary Tables~\ref{tab:1} to \ref{tab:5}. We also choose a typical link to show the key rate performance for over a week in Supplementary Figure~\ref{Fig:keyratedetail}.


\subsection*{Supplementary Note 2: Key Rate Calculation}\label{app:keyrate}
We take a typical link UB-23 to TR-3 in Supplementary Table~\ref{tab:1} as an example to show how to calculate the finite-size key rate with experimental data. We recall the BB84 key rate formula given in the main text,
\begin{equation}\label{eq:keyrateBB84supp}
r=-fQ_\mu H(E_\mu)+Y_1\mu e^{-\mu}[1-H(e_1^p)].
\end{equation}
In our experiment, we apply vacuum and weak decoy state method to estimate the lower bound of $Y_1$ and the upper bound of $e_1$, which are denoted by $Y_1^L$ and $e_1^{U}$, respectively. These bounds are given by \cite{Ma05}
\begin{equation}\label{eq:decoy}
\begin{aligned}
Y_1^L & = \frac{\mu}{\mu \nu - \nu^2} \left(Q_\nu e^\nu -Q_\mu e^\mu \frac{\nu^2}{\mu^2} - \frac{\mu^2-\nu^2}{\mu^2} Y_0 \right) \\
e_1^{U} & = \frac{E_\nu Q_\nu e^\nu -e_0 Y_0}{Y_1^L \nu},
\end{aligned}
\end{equation}
where $\mu$ and $\nu$ are intensities of signal and weak decoy states, respectively, $Q_\mu$ and $Q_\nu$ are gains of signal and weak decoy states, respectively, $Y_0$ equals to the dark count rate $p_d$, and $e_0$ is the error rate of dark counts and equals to $0.5$. Considering the symmetry of $Z$ and $X$ basis, we do not denote the basis information for the parameters above. One can calculate the bounds above in each basis.
Note that $e_1^{U}$ is the the upper bound of the bit error rate of the single photon component in $X$ or $Z$ basis. Due to the basis-independence of single photon component, we have $e_1^{pz} = e_1^{bx}$ and $e_1^{px}=e_1^{bz}$. Then we can directly substitute $e_1^{U}$ into Eq.~\eqref{eq:keyrateBB84supp} as the upper bound of the phase error rate in asymptotic case.

In finite-size case, we need to consider the deviation between $e_1^{pz}$ and $e_1^{bx}$ (or between $e_1^{px}$ and $e_1^{bz}$), which is given by the random sampling method \cite{PhysRevA.81.012318},
\begin{equation}\label{eq:deviation}
\begin{aligned}
P_{\theta x} & = \mathrm{Prob}\{e_1^{pz} \geq e_1^{bx}+\theta\} \\
& < \frac{\sqrt{n_x +n_z}}{\sqrt{e_1^{bx}(1-e_1^{bx})}} 2^{-(n_x+n_z)\xi(\theta)},
\end{aligned}
\end{equation}
and
\begin{equation}
\xi(\theta) = h(e_1^{bx}+\theta -q_x \theta) - q_x h(e_1^{bx}) - (1-q_x)h(e_1^{bx}+\theta),
\end{equation}
where $n_x$ and $n_z$ are the number of sifted key bits in $X$ and $Z$ basis, respectively, and $q_x = n_x/(n_x+n_z)$.

We also need to consider the statistical fluctuation in the observed experimental data. Here we assume Gaussian fluctuations and set the fluctuations to be $\delta = 10$ standard deviations, then we have the following bounds for an arbitrary variable $\chi$
\begin{equation}\label{eq:Gaussiandeviation}
\begin{aligned}
\chi^L &  = \chi - \frac{\delta \sqrt{\chi}}{\sqrt{N}}   \\
\chi^U &  = \chi + \frac{\delta \sqrt{\chi}}{\sqrt{N}},   \\
\end{aligned}
\end{equation}
where $\chi$ can be the gain, $Q_\mu$ and $Q_\nu$, quantum bit error rate in each basis, $E_\mu Q_\mu$ and $E_\nu Q_\nu$, the successful detections for single photon components, $M_1^{zsL}$ and $M_1^{xL}$. The last two quantities cannot be directly observed in experiment. They are defined as
\begin{equation}
\begin{aligned}
M_1^{zsL} & = n_z Y_1^L q^s \mu e^{-\mu} \\
M_1^{xL} & = n_x Y_1^L (q^s \mu e^{-\mu} + q^d \nu e^{-\nu})
\end{aligned}
\end{equation}
where $n_z$ and $n_x$ are the number of $Z$ and $X$ basis states after sifting, $q^s$ and $q^d$ are the probabilities of choosing signal and weak decoy states, respectively. In our case $q^s=0.75$ and $q^d=0.125$.

We use the following substitutions in Eqs.\eqref{eq:decoy} and \eqref{eq:deviation} after considering the statistical fluctuation, $Q_\mu \rightarrow Q_\mu^U$, $Q_\nu \rightarrow Q_\nu^L$, $e_1^{bx}\rightarrow e_1^{U}$, $e_1^{pz} \rightarrow e_1^{psz}$ $n_x \rightarrow M_1^{xL}$, and $n_z \rightarrow M_1^{zsL}$. The finite-key length in $Z$ basis is given by
\begin{equation}
\begin{aligned}
K^z \geq M_1^{zsL}\left[1-H(e_1^{psz})\right] - M^{sz} f H(E_\mu) - \Delta,
\end{aligned}
\end{equation}
where $\Delta$ is the total key consumption in the post-processing \cite{PhysRevA.81.012318} and $M^{sz}$ is the number of successful detections of $Z$-basis signal states. If we take the failure probability in each step as $10^{-10}$, $\Delta$ will be at the order of $10^2$ bits. The finite-key length $K^x$ in $X$ basis can be calculated similarly. In our case, we use both basis to generate keys, the total finite-key length is $K^{tot} = K^x+K^z$. The key rate is given by $r_0 K^{tot}/N$, where $r_0$ is the repetition rate of the source.

The values of the parameters of the link UB-23 to TR-3 in the key rate calculation are listed in Supplementary Table~\ref{tab:parameter}, with which we can calculate the finite-size key rate in Supplementary Table~\ref{tab:1}. In the future, we will update the key rate calculation module with the Chernoff-Hoeffding method in our system.

\subsection*{Supplementary Figures and Tables}
\begin{figure}[hbt]
\centering
\includegraphics[width=10cm]{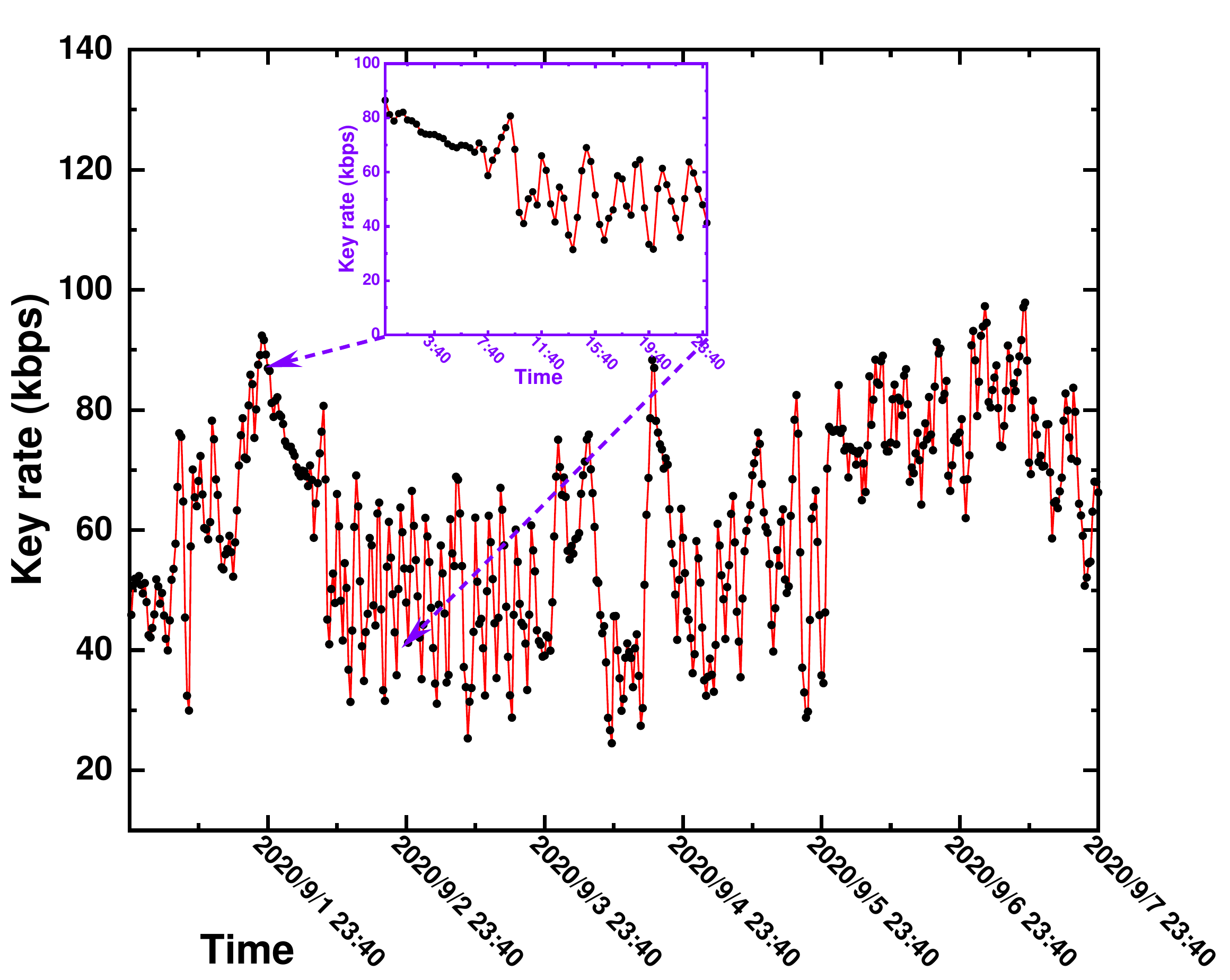}
\caption{Key rate performance of a typical link for over a week. The key rates are collected over a week from Sept. 1 to Sept 7, 2020, are averaged every 20 minutes. The inset shows the detailed data within 24 hours on Sept. 2, 2020. The transmission loss for this link is $4.1$ dB.}
\label{Fig:keyratedetail}
\end{figure}
\begin{table*}[ht!]
\centering
\caption{Feedback calibration time of the TR2-TR3 link over 24 hours on April 23, 2021.}
\begin{tabular}{|c|c|}
\hline
\textbf{Polarization Calibration Time} & \textbf{Calibration Interval}\\
(hour:minute:second) & (hour:minute:second) \\
\hline
$0:19:43$ & ---\\
$11:19:18$ & $10:59:35$\\
$11:45:32$	& $0:26:14$ \\
$12:43:02$	&$0:57:30$ \\
$13:41:56$	&$0:58:54$ \\
$14:08:35$	&$0:26:39$ \\
$15:20:13$	&$1:11:38$ \\
$15:20:53$	&$0:00:40$ \\
$15:35:13$	&$0:14:20$ \\
$15:59:44$	&$0:24:31$ \\
$16:32:35$	&$0:32:51$ \\
$16:59:30$	&$0:26:55$ \\
$19:40:18$	&$2:40:48$\\
$21:02:01$	&$1:21:43$ \\
$23:34:36$	&$2:32:35$ \\
\hline
\end{tabular}
\end{table*}

\begin{table*}[ht!]
\centering
\caption{Key rates (unit: $\rm kbps$) among nodes in the QuantumCTek subnetwork. TR: Trusted Relay; UA: Type-A users that connected to an optical switch; UB: Type-B users that connected to a trusted relay.}
\begin{tabular}{|c|c|c||c|c|c|}
\hline
\textbf{Transmitter} & \textbf{Receiver} & \textbf{Key rate}& \textbf{Transmitter} & \textbf{Receiver} & \textbf{Key rate}\tabularnewline
\hline
\hline
\multirow{2}{*}{TR-3} & TR-1 & 19.8 & \multirow{6}{*}{UA-10} & UA-8 & 20\tabularnewline
\cline{2-3} \cline{5-6}
 & TR-2 & 61.4 &  & UA-9 & 32\tabularnewline
\cline{1-3} \cline{5-6}
UB-18 & \multirow{10}{*}{TR-3} & 7 &  & UA-11 & 30\tabularnewline
\cline{1-1} \cline{3-3} \cline{5-6}
UB-19 &  & 12 &  & UA-12 & 37\tabularnewline
\cline{1-1} \cline{3-3} \cline{5-6}
UB-20 &  & 10 &  & UA-13 & 32\tabularnewline
\cline{1-1} \cline{3-3} \cline{5-6}
UB-21 &  & 9 &  & TR-3 & 8\tabularnewline
\cline{1-1} \cline{3-6}
UB-22 &  & 10 & \multirow{6}{*}{UA-11} & UA-8 & 21\tabularnewline
\cline{1-1} \cline{3-3} \cline{5-6}
UB-23 &  & 18 &  & UA-9 & 20\tabularnewline
\cline{1-1} \cline{3-3} \cline{5-6}
UB-24 &  & 8 &  & UA-10 & 30\tabularnewline
\cline{1-1} \cline{3-3} \cline{5-6}
UB-25 &  & 11 &  & UA-12 & 19\tabularnewline
\cline{1-1} \cline{3-3} \cline{5-6}
UB-26 &  & 18 &  & UA-13 & 26\tabularnewline
\cline{1-1} \cline{3-3} \cline{5-6}
UB-27 &  & 8 &  & TR-3 & 9\tabularnewline
\hline
\multirow{6}{*}{UA-8} & UA-9 & 19 & \multirow{6}{*}{UA-12} & UA-8 & 20\tabularnewline
\cline{2-3} \cline{5-6}
 & UA-10 & 20 &  & UA-9 & 24\tabularnewline
\cline{2-3} \cline{5-6}
 & UA-11 & 21 &  & UA-10 & 37\tabularnewline
\cline{2-3} \cline{5-6}
 & UA-12 & 20 &  & UA-11 & 19\tabularnewline
\cline{2-3} \cline{5-6}
 & UA-13 & 22 &  & UA-13 & 33\tabularnewline
\cline{2-3} \cline{5-6}
 & TR-3 & 10 &  & TR-3 & 6\tabularnewline
\hline
\multirow{6}{*}{UA-9} & UA-8 & 19 & \multirow{6}{*}{UA-13} & UA-8 & 22\tabularnewline
\cline{2-3} \cline{5-6}
 & UA-10 & 32 &  & UA-9 & 25\tabularnewline
\cline{2-3} \cline{5-6}
 & UA-11 & 20 &  & UA-10 & 32\tabularnewline
\cline{2-3} \cline{5-6}
 & UA-12 & 24 &  & UA-11 & 26\tabularnewline
\cline{2-3} \cline{5-6}
 & UA-13 & 25 &  & UA-12 & 33\tabularnewline
\cline{2-3} \cline{5-6}
 & TR-3 & 10 &  & TR-3 & 28\tabularnewline
\hline
\end{tabular}
\label{tab:1}
\end{table*}

\begin{table*}[ht!]
\centering
\caption{Key rates (unit: $\rm kbps$) among nodes in the city library subnetwork. TR: Trusted Relay; UA: Type-A users that connected to an optical switch; UB: Type-B users that connected to a trusted relay.}
\begin{tabular}{|c|c|c||c|c|c|}
\hline
\textbf{Transmitter} & \textbf{Receiver} & \textbf{Key rate} & \textbf{Transmitter} & \textbf{Receiver} & \textbf{Key rate}\tabularnewline
\hline
\hline
\multirow{2}{*}{TR-2} & TR-1 & 23.7 & \multirow{5}{*}{UA-3} & UA-1 & 18\tabularnewline
\cline{2-3} \cline{5-6}
 & TR-3 & 54.4 &  & UA-2 & 24\tabularnewline
\cline{1-3} \cline{5-6}
UB-13 & \multirow{5}{*}{TR-2} & 13 &  & UA-4 & 19\tabularnewline
\cline{1-1} \cline{3-3} \cline{5-6}
UB-14 &  & 17 &  & UA-5 & 28\tabularnewline
\cline{1-1} \cline{3-3} \cline{5-6}
UB-15 &  & 12 &  & TR-2 & 17\tabularnewline
\cline{1-1} \cline{3-6}
UB-16 &  & 16 & \multirow{5}{*}{UA-4} & UA-1 & 11\tabularnewline
\cline{1-1} \cline{3-3} \cline{5-6}
UB-17 &  & 21 &  & UA-2 & 11\tabularnewline
\cline{1-3} \cline{5-6}
\multirow{5}{*}{UA-1} & UA-2 & 17 &  & UA-3 & 19\tabularnewline
\cline{2-3} \cline{5-6}
 & UA-3 & 18 &  & UA-5 & 25\tabularnewline
\cline{2-3} \cline{5-6}
 & UA-4 & 11 &  & TR-2 & 27\tabularnewline
\cline{2-6}
 & UA-5 & 10 & \multirow{5}{*}{UA-5} & UA-1 & 10\tabularnewline
\cline{2-3} \cline{5-6}
 & TR-2 & 30 &  & UA-2 & 21\tabularnewline
\cline{1-3} \cline{5-6}
\multirow{5}{*}{UA-2} & UA-1 & 17 &  & UA-3 & 28\tabularnewline
\cline{2-3} \cline{5-6}
 & UA-3 & 24 &  & UA-4 & 25\tabularnewline
\cline{2-3} \cline{5-6}
 & UA-4 & 11 &  & TR-2 & 16\tabularnewline
\cline{2-6}
 & UA-5 & 21 & \multirow{2}{*}{UA-6} & UA-7 & 11\tabularnewline
\cline{2-3} \cline{5-6}
 & TR-2 & 11 &  & TR-2 & 10\tabularnewline
\hline
UA-7 & UA-6 & 11 & UA-7 & TR-2 & 13\tabularnewline
\hline
\end{tabular}
\label{tab:2}
\end{table*}

\begin{table*}
\centering
\caption{Key rates (unit: $\rm kbps$) among nodes in the USTC subnetwork. TR: Trusted Relay; UA: Type-A users that connected to an optical switch; UB: Type-B users that connected to a trusted relay.}
\begin{tabular}{|c|c|c|}
\hline
\textbf{Transmitter} & \textbf{Receiver} & \textbf{Key rate}\tabularnewline
\hline
\hline
\multirow{2}{*}{TR-1} & TR-2 & 7.8\tabularnewline
\cline{2-3}
 & TR-3 & 49.5 \tabularnewline
\hline
UB-1 & \multirow{12}{*}{TR-1} & 6\tabularnewline
\cline{1-1} \cline{3-3}
UB-2 &  & 16\tabularnewline
\cline{1-1} \cline{3-3}
UB-3 &  & 17\tabularnewline
\cline{1-1} \cline{3-3}
UB-4 &  & 16\tabularnewline
\cline{1-1} \cline{3-3}
UB-5 &  & 9\tabularnewline
\cline{1-1} \cline{3-3}
UB-6 &  & 17\tabularnewline
\cline{1-1} \cline{3-3}
UB-7 &  & 9\tabularnewline
\cline{1-1} \cline{3-3}
UB-8 &  & 17\tabularnewline
\cline{1-1} \cline{3-3}
UB-9 &  & 6\tabularnewline
\cline{1-1} \cline{3-3}
UB-10 &  & 8\tabularnewline
\cline{1-1} \cline{3-3}
UB-11 &  & 14\tabularnewline
\cline{1-1} \cline{3-3}
UB-12 &  & 12\tabularnewline
\hline
\end{tabular}
\label{tab:3}
\end{table*}

\begin{table*}[]
\centering
\caption{Detailed key rates (unit: $\rm kbps$) in the robustness test among the trusted relay and 5 user nodes in city library subnetwork over $31$ months (December 2017 to June 2020). For this robustness test, we have upgraded the system repetition rate from 20 MHz to 40 MHz. TR: Trusted Relay; UB: Type-B users that connected to a trusted relay. The numbers correspond to the key rates with units of kbps.}
\begin{tabular}{|c|c|c|c|c|c|}
\hline
\multirow{2}{*}{Time (month)} & \multicolumn{5}{c|}{TR-2}                            \\ \cline{2-6}
                              & UB-13 & UB-16 & UB-14 & UB-15 & UB-17 \\ \hline
1                             & 26.1 & 8.3  & 11.4 & 35.6 & 8.5    \\ \hline
2                             & 32.9 & 4.8  & 18.5 & 35.0 & 7.1    \\ \hline
3                             & 39.1 & 6.4  & 22.8 & 45.5 & 7.6    \\ \hline
4                             & 38.6 & 8.8  & 20.2 & 47.8 & 6.4   \\ \hline
5                             & 44.8 & 10.7 & 23.9 & 48.4 & 8.7    \\ \hline
6                             & 47.4 & 14.0 & 23.8 & 56.8 & 11.2   \\ \hline
7                             & 37.6 & 10.9 & 19.0 & 50.6 & 13.7  \\ \hline
8                             & 46.8 & 11.3 & 13.7 & 53.4 & 12.4  \\ \hline
9                             & 45.0 & 10.8 & 12.3 & 54.5 & 9.0   \\ \hline
10                            & 58.4 & 11.4 & 10.4 & 55.1 & 7.0   \\ \hline
11                            & 49.4 & 9.5  & 11.4 & 56.1 & 9.5   \\ \hline
12                            & 49.0 & 9.8  & 5.6  & 56.8 & 8.5    \\ \hline
13                            & 48.2 & 9.4  & 7.2  & 15.1 & 8.5    \\ \hline
14                            & 71.4 & 8.0  & 8.1  & 27.3 & 13.7   \\ \hline
15                            & 53.4 & 9.1  & 9.8  & 21.7 & 12.2   \\ \hline
16                            & 57.1 & 22.5 & 8.0  & 56.1 & 11.2   \\ \hline
17                            & 55.9 & 26.4 & 3.4  & 56.1 & 13.5  \\ \hline
18                            & 53.5 & 9.7  & 5.3  & 60.4 & 12.7   \\ \hline
19                            & 33.9 & 14.3 & 5.1  & 53.9 & 11.6   \\ \hline
20                            & 45.3 & 18.5 & 4.2  & 58.4 & 9.9    \\ \hline
21                            & 50.4 & 17.5 & 6.1  & 59.8 & 10.3   \\ \hline
22                            & 45.4 & 16.4 & 7.3  & 58.1 & 12.8  \\ \hline
23                            & 31.4 & 15.6 & 6.9  & 57.4 & 14.6   \\ \hline
24                            & 44.7 & 13.7 & 9.0  & 57.8 & 9.9    \\ \hline
25                            & 55.1 & 14.3 & 12.7 & 50.5 & 9.9    \\ \hline
26                            & 38.2 & 17.2 & 14.7 & 54.5 & 12.7  \\ \hline
27                            & 43.7 & 17.1 & 16.8 & 56.3 & 13.2  \\ \hline
28                            & 39.5 & 17.7 & 15.2 & 56.1 & 12.9   \\ \hline
29                            & 37.3 & 18.8 & 16.5 & 55.1 & 13.3  \\ \hline
30                            & 35.0 & 16.9 & 13.6 & 52.0 & 10.5   \\ \hline
31                            & 35.0 & 16.6 & 14.0 & 50.1 & 13.9   \\ \hline
\end{tabular}
\label{tab:4}
\end{table*}
\begin{table*}[ht!]
\centering
\caption{Detailed key rates (unit: $\rm kbps$) in the robustness test among the trusted relay and 6 user nodes in QuantumCTek subnetwork over $31$ months (December 2017 to June 2020). For this robustness test, we have upgraded the system repetition rate from 20 MHz to 40 MHz. TR: Trusted Relay; UB: Type-B users that connected to a trusted relay. The numbers correspond to the key rates with units of kbps.}
\begin{tabular}{|c|c|c|c|c|c|c|}
\hline
\multirow{2}{*}{Time (month)}         & \multicolumn{6}{c|}{TR-3}                     \\ \cline{2-7}
                              & UB-19 & UB-20 & UB-21 & UB-22 & UB-23 & UB-24 \\ \hline
1                               & 7.7   & 34.7  & 19.3  & 29.9  & 6.7   & 8.5   \\ \hline
2                              & 6.0   & 30.7  & 22.2  & 27.1  & 16.8  & 41.6  \\ \hline
3                              & 9.9   & 31.1  & 22.2  & 25.1  & 27.1  & 47.5  \\ \hline
4                               & 8.7   & 30.7  & 19.4  & 26.3  & 26.5  & 46.9  \\ \hline
5                              & 8.3   & 31.1  & 21.5  & 28.4  & 26.0  & 45.3  \\ \hline
6                              & 10.3  & 30.5  & 22.7  & 30.5  & 25.9  & 42.8  \\ \hline
7                              & 5.9   & 24.8  & 19.9  & 35.9  & 23.9  & 40.0  \\ \hline
8                             & 8.7   & 26.0  & 18.7  & 51.2  & 17.6  & 52.3  \\ \hline
9                              & 9.7   & 26.8  & 20.1  & 51.6  & 18.1  & 44.0  \\ \hline
10                             & 8.4   & 24.7  & 22.1  & 44.5  & 17.9  & 35.8  \\ \hline
11                             & 8.3   & 32.0  & 21.1  & 53.8  & 18.9  & 43.6  \\ \hline
12                             & 10.0  & 30.0  & 19.4  & 54.7  & 23.1  & 44.9  \\ \hline
13                             & 8.5   & 27.4  & 16.6  & 47.9  & 21.6  & 54.8  \\ \hline
14                            & 10.0  & 29.4  & 16.7  & 51.7  & 21.3  & 42.3  \\ \hline
15                             & 8.3   & 23.3  & 22.8  & 55.1  & 27.5  & 53.1  \\ \hline
16                             & 7.0   & 28.7  & 21.4  & 39.1  & 21.7  & 45.9  \\ \hline
17                             & 7.1   & 29.6  & 22.7  & 38.3  & 18.3  & 37.4  \\ \hline
18                             & 4.6   & 29.6  & 21.0  & 41.8  & 16.2  & 41.5  \\ \hline
19                            & 6.4   & 28.0  & 19.7  & 35.4  & 17.5  & 45.1  \\ \hline
20                              & 3.9   & 32.6  & 20.1  & 46.8  & 18.2  & 43.9  \\ \hline
21                             & 6.3   & 32.6  & 20.3  & 45.2  & 18.0  & 40.5  \\ \hline
22                             & 7.2   & 27.9  & 17.1  & 38.3  & 14.0  & 37.0  \\ \hline
23                             & 9.0   & 22.6  & 15.1  & 40.3  & 16.5  & 37.3  \\ \hline
24                              & 7.3   & 24.7  & 16.5  & 38.0  & 21.3  & 42.5  \\ \hline
25                             & 7.4   & 28.4  & 15.0  & 35.0  & 20.9  & 39.5  \\ \hline
26                             & 7.0   & 26.6  & 14.1  & 36.5  & 23.8  & 41.9  \\ \hline
27                             & 7.7   & 24.3  & 15.3  & 38.2  & 24.6  & 41.6  \\ \hline
28                            & 8.6   & 20.4  & 13.4  & 40.7  & 24.2  & 40.1  \\ \hline
29                             & 8.3   & 22.8  & 14.0  & 36.6  & 28.8  & 36.8  \\ \hline
30                             & 7.3   & 13.5  & 16.2  & 37.8  & 24.9  & 32.9  \\ \hline
31                            & 7.2   & 19.6  & 16.6  & 42.9  & 21.4  & 26.6  \\ \hline
\end{tabular}
\label{tab:5}
\end{table*}

\begin{table*}[ht!]
\centering
\caption{Parameters in key rate calculation}
\begin{tabular}{ccccccc}
\hline
\hline
$p_d$  & $Q_\mu$ & $Q_{\nu}$ & $\mu$ & $\nu$ & $f$ & $E_\mu$\\
\hline
$10^{-6}$ & $0.067$ & $0.022$  &  $0.6$ & $0.2$ & $1.5$ & $0.01$\\
\hline
\hline
$E_\nu$ & $N$ & $\epsilon_{ph}$   & $q^s$ &$q^d$ & $r_0$\\
\hline
$0.024$ & $6.3 \times 10^7$ & $10^{-10}$  & $0.75$ & $0.125$ & $40$MHz \\
\hline
\hline
\end{tabular}
\label{tab:parameter}
\end{table*}

\bibliography{BibNet}

\end{document}